\documentclass[12pt]{article}

\usepackage[T1]{fontenc}
\usepackage[utf8]{inputenc}
\usepackage{times}
\usepackage{geometry}
\usepackage{setspace}
\usepackage{physics}
\usepackage{graphicx}
\usepackage{amssymb}
\usepackage{amsmath}
\usepackage{dsfont}
\usepackage{bm}
\usepackage{mathrsfs}
\usepackage{bbold}
\usepackage{xcolor}
\usepackage{amsthm}
\usepackage{nameref}
\usepackage{hyperref}
\usepackage{enumitem}

\hypersetup{
  colorlinks=true,
  linkcolor=blue,
  citecolor=blue,
  urlcolor=blue
}

\newtheoremstyle{natural}%
{\topsep}{\topsep}%
{\normalfont}{}%
{\normalfont}{.}{5pt}%
{}
\theoremstyle{natural}

\begin{document}
\thispagestyle{empty}

\begin{center}
{\Large \textbf{The Unruh Effect in Relativistic Fluids}}\\[1em]

\textbf{Eren Erberk Erkul}$^{|\star\rangle,|\dagger\rangle}$\\
{\small
$^{|\star\rangle}$Institute of Science and Technology Austria (ISTA), 3400 Klosterneuburg, Austria\\
$^{|\dagger\rangle}$Department of Physics \& Department of Electrical and Electronics Engineering,\\
Middle East Technical University (METU), Ankara 06800, Turkey
}\\[0.5em]
\textbf{Dated:} October 9, 2025
\end{center}

\begin{center}
\textbf{Abstract}
\end{center}

\begin{center}
\begin{minipage}{0.87\textwidth}
\normalsize
We identify the relativistic-fluid counterpart of the Unruh effect, in which a comoving probe measures a \emph{Thermodynamic Unruh temperature}. Frame changes in first-order hydrodynamics are recast as a local, time-dependent hyperbolic rotation in a Rindler-style state space where the instantaneous map between frames is the \emph{Thermodynamic Boost} and its proper-time variation defines the \emph{Thermodynamic Acceleration}, which results in an Unruh-like thermal spectrum. To leading order, the \emph{Thermodynamic Unruh temperature} is frame-independent and universal across out-of-equilibrium relativistic fluid descriptions, from Israel--Stewart to modern theories.

\end{minipage}
\end{center}

\begin{center}
\vspace{8\baselineskip}
{\bfseries\itshape To the stubborn light.}
\vspace{0.75\baselineskip}
\end{center}

\clearpage
\pagenumbering{arabic}

\noindent\textbf{Conventions.} We adopt the metric signature $(-,+,+,+)$ and set $c=\hbar=k_B=1$ except where we display them explicitly. The fluid four-velocity satisfies $u^\mu u_\mu=-1$. The projector is $\Delta^{\mu\nu}=g^{\mu\nu}+u^\mu u^\nu$. Angle brackets denote transverse (and, for rank-2, symmetric traceless) projections: $A^{\langle\mu\rangle}\equiv \Delta^\mu{}_\nu A^\nu$, $B^{\langle\mu\nu\rangle}\equiv \Delta^{\mu\nu}{}_{\alpha\beta}B^{\alpha\beta}$.

\section{Introduction}

The Unruh effect establishes a connection between geometry and thermodynamics where an observer accelerating uniformly through the Minkowski vacuum experiences an additional thermal bath with a temperature proportional to their proper acceleration\cite{Unruh1976,Crispino2008RMP}. This phenomenon, along with the closely related Hawking effect \cite{Hawking1975}, suggests that the particle content of a quantum field theory is observer-dependent.

Observation of the Unruh effect requires accelerations far beyond those found in the laboratory, and direct verification is unattainable with existing setups. Nevertheless, there are experimental efforts and theoretical proposals aimed at verifying it through analogous classical systems, ranging from water-wave horizons to optical media, which seek to demonstrate the universal ingredients of these quantum vacuum effects.\cite{Vallisneri1998,Leonhardt2018}

In this paper, we aim to go beyond these analogues and to show that an exact counterpart of the Unruh effect emerges intrinsically within the structure of relativistic hydrodynamics. The first-order relativistic hydrodynamics suffers from a fundamental ambiguity in defining the local fluid rest frame (the four-velocity $u^\mu$). For conceptual clarity and historical alignment (see Appendix~\ref{sec:relfluid}), the two canonical choices are the Eckart frame,\cite{Eckart1940}, which follows the flow of conserved  particles, and the Landau--Lifshitz frame,\cite{Landau1959}, which follows the flow of energy. Out of equilibrium, these frames differ (see also Refs.~\cite{GerochLindblom1990,RomatschkeRomatschke2019,RezzollaZanotti2013}). A fluid element described as having only energy density in the Landau frame appears to possess both energy density and a heat flux when described in the Eckart frame.

\newpage

We propose a novel geometric interpretation of this ambiguity. Frame changes can be recast as a local, time-dependent hyperbolic rotation in an abstract Rindler-style  \emph{thermodynamic heat-plane.} The instantaneous map between frames is a Lorentz-like boost we call \emph{Thermodynamic Boost} whose parameters are fixed by the local state (e.g., $q/(\varepsilon+p)$) and generally vary along the worldline.

When the fluid evolves dynamically, the required boost parameters naturally change over time. This proper-time variation is described by the time variation of the "Thermodynamic Boost," which we call "Thermodynamic Acceleration." Then, continuing with the recipe of the standard Unruh effect, we hypothesize that this thermodynamic acceleration leads to a new spectral signature, a Thermodynamic Unruh temperature $T_{\rm th}$ that is perceptible by a comoving probe. To make a more formal derivation, using the Israel--Stewart theory \cite{Israel1976,IsraelStewart1979}, we derived an expression for $T_{\rm th}$ in terms of  hydrodynamic variables and transport coefficients. Finally, we demonstrate that our arguments extend to modern relativistic fluid theories; these theories contribute only through corrections to transport coefficients, without invalidating the basic mechanism. Hence, the signal is present regardless of frame choice (including general first-order BDNK frames) and is not an artifact of deficiencies in the Eckart or Landau frames.
\newpage

\section{Geometric Frame Dynamics}
\label{sec:thermo-boosts}

The ambiguity between the \emph{Eckart} and \emph{Landau--Lifshitz} frames\, is a well-known feature of relativistic hydrodynamics. We propose viewing frame changes as a hyperbolic rotation in an abstract thermodynamic heat-plane; the instantaneous map between a working frame and the energy frame is a Lorentz-like Thermodynamic Boost.

Let $u^\mu_{E}$ and $u^\mu_{L}$ denote the four-velocities of the Eckart and Landau frames, respectively. In the Eckart frame, the energy density is $\varepsilon_{E}$ and the magnitude of the heat flux is $q_{E}$. In the Landau frame, the rest energy density is $\varepsilon_{L}$, and the heat flux $q_L$ vanishes by definition. The instantaneous points in frames are related by a Lorentz-like transformation, where the relative velocity $v$ and Lorentz factor $\gamma=(1-v^{2})^{-1/2}$ have their thermodynamic interpretations. We denote enthalpy density as $w=\varepsilon+p$.

To illustrate the hyperbolic nature of this transformation, consider the energy--momentum flow vector $J^\mu\equiv -\,T^\mu{}_{\nu}u^\nu$. In the Eckart rest frame, restricted to the direction of heat flow, this vector is
\begin{equation}
J^\mu_E = (\varepsilon_E,\, q_E).
\label{eq:JmuE}
\end{equation}
Its invariant squared length,
\begin{equation}
J^\mu J_\mu = -\varepsilon_E^2 + q_E^2,
\label{eq:J-invariant}
\end{equation}
is frame independent. A Lorentz transformation to a frame moving at velocity $v$ relative to Eckart mixes the components:
\begin{align}
\varepsilon_L &= \gamma(\varepsilon_E - v\, q_E), \label{eq:landau-boost-eps}\\
q_L &= \gamma(q_E - v\, \varepsilon_E).
\label{eq:landau-boost-q}
\end{align}
This transformation highlights how energy density and heat flux are observer-dependent concepts that interconvert under a change of hydrodynamic frame.

To find the specific velocity $v$ that reaches the Landau frame, we impose the defining condition $q_L=0$. If we momentarily neglect pressure contributions and define the velocity simply by the ratio of energy flux to energy density, we find
\begin{equation}
v = \frac{q_E}{\varepsilon_E}.
\label{eq:v-simple}
\end{equation}
This leads to the Lorentz factor
\begin{equation}
\gamma = \frac{1}{\sqrt{1-v^2}}
= \frac{\varepsilon_E}{\sqrt{\varepsilon_E^2 - q_E^2}}.
\label{eq:gamma}
\end{equation}
Substituting back into Eq.~\eqref{eq:landau-boost-eps} yields
\begin{align}
\varepsilon_L
&= \left( \frac{\varepsilon_E}{\sqrt{\varepsilon_E^2 - q_E^2}} \right)
\left( \varepsilon_E - \frac{q_E^2}{\varepsilon_E} \right) \nonumber\\
&= \sqrt{\varepsilon_E^2 - q_E^2}.
\label{eq:epsL-sqrt}
\end{align}
Squaring this result reveals an illustrative hyperbolic constraint on the fluid state space,
\begin{equation}
\varepsilon_E^2 - q_E^2 = \varepsilon_L^2.
\label{eq:hyperbolic-constraint}
\end{equation}
confirming the geometric structure of a boost.

Hence, to generalize to cases with pressure, we replace $\varepsilon_E$ with the enthalpy density
$w_E \equiv \varepsilon_E + p_E$ in Eq.~\eqref{eq:v-simple}, the leading-order boost velocity that maps a working frame to the energy frame is
\begin{equation}
v \;=\; \frac{q_E}{\varepsilon_E + p_E} \;+\; \mathcal{O}\!\left(\frac{q^3}{(\varepsilon+p)^3}\right).
\label{eq:v-correct}
\end{equation}

At any instant, for small heat flux, the local boost that maps a working frame (e.g., Eckart) to the energy frame (Landau, $q=0$) admits the first-order approximation
\begin{equation}
u^\mu_{(L)} = u^\mu_{(E)} + \frac{q^\mu_{(E)}}{\varepsilon_E+p_E} + \mathcal{O}(q^2).
\label{eq:uL-linear-2}
\end{equation}
The accuracy of this approximation is verified by checking the norm of $u^\mu_{(L)}$, which must remain $-1$:
\begin{align}
u^\mu_L u_{\mu L}
&= \left(u^\mu_E + \frac{q^\mu_E}{\varepsilon_E+p_E}\right)
\left(u_{\mu E} + \frac{q_{\mu E}}{\varepsilon_E+p_E}\right) \nonumber\\
&= -1 + 0 + \frac{q_E^2}{(\varepsilon_E+p_E)^2}
= -1 + \mathcal{O}(q^2).
\label{eq:uL-norm}
\end{align}
In effective field theory (EFT) treatments focused on linear order, this second-order correction can be ignored. Imposing the normalization condition of the four-velocity under this thermodynamic boost provides a geometric justification for the linear-order analysis.

\textbf{Geometric 1+1+2 split.}
A more rigorous realization of the boost structure, which naturally incorporates pressure anisotropies, is provided by decomposing the stress--energy tensor $T^{\mu\nu}$ (see, e.g., Refs.~\cite{Kovtun2012}). We define a basis aligned with the heat flow. Let $u^\mu$ be the working frame (e.g., Eckart) and $\hat q^{\mu}\equiv q^{\mu}/|q|$ the normalized heat direction, with $u\!\cdot\!\hat q=0$ and $\hat q^2=1$. Define the longitudinal projector $P_{\parallel}^{\mu\nu}=\hat q^{\mu}\hat q^{\nu}$ and the transverse projector $P_{\perp}^{\mu\nu}=\Delta^{\mu\nu}-\hat q^{\mu}\hat q^{\nu}$. The stress-energy tensor decomposes as:
\begin{equation}
\begin{aligned}
T^{\mu\nu} &=
\varepsilon\, u^\mu u^\nu
+ p_{\parallel}\, \hat q^\mu \hat q^\nu
+ p_{\perp}\, P_{\perp}^{\mu\nu}
+ q \left(u^\mu \hat q^\nu + u^\nu \hat q^\mu\right) \\
&\quad + 2 \hat q^{(\mu}\pi_{\perp}^{\nu)}
+ \tilde\pi_{\perp}^{\mu\nu},
\end{aligned}
\label{eq:Tmunu_decomp}
\end{equation}
with $p_{\parallel}=p+\Pi+\pi_{\parallel}$ and $p_{\perp}=p+\Pi-\tfrac{1}{2}\pi_{\parallel}$. Here $\Pi$ is the bulk viscous pressure, $\pi_{\parallel}$ is the longitudinal shear component, and the remaining terms represent transverse shear stresses in the 2-surface orthogonal to $\{u,\hat q\}$.

\begin{figure}[ht]
\centering
\includegraphics[width=0.5\textwidth]{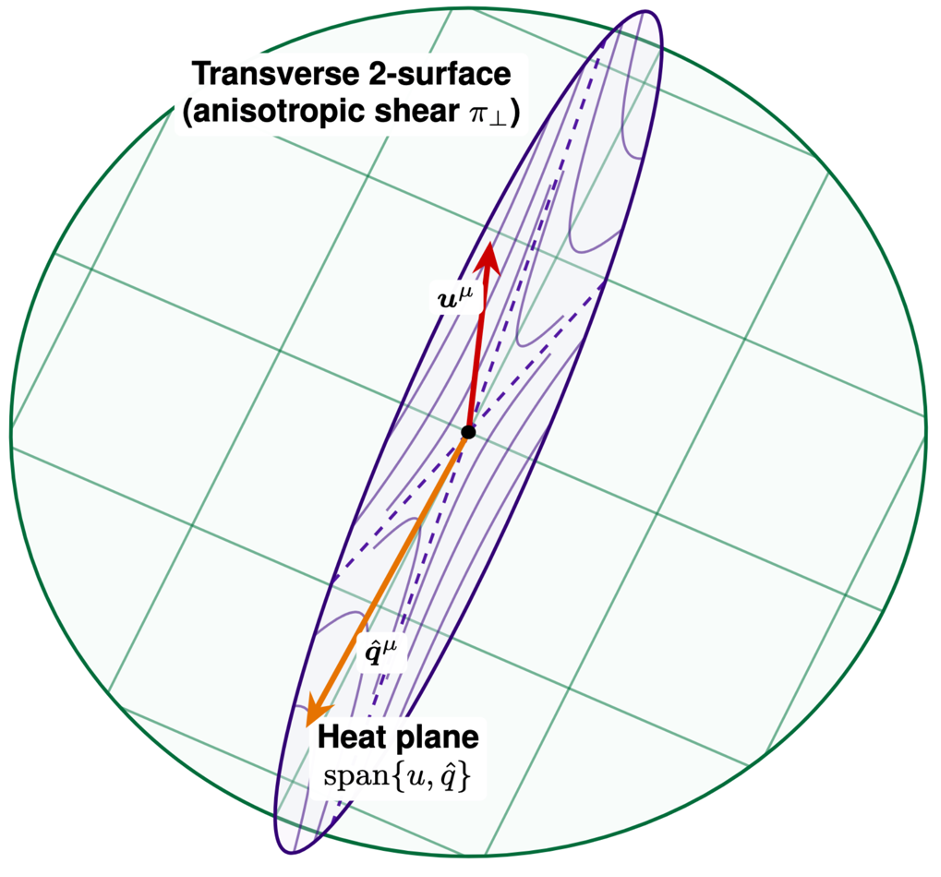}
\caption{\textbf{The space of the $1+1+2$ decomposition.} The heat plane $\mathrm{span}\{u,\hat q\}$ carrying the hyperbolic block and the transverse 2-surface carrying anisotropic shear.}
\label{fig:sketch1p1p2}
\end{figure}

Restricting our attention to the ``heat plane'' $\mathrm{span}\{u,\hat q\}$ (see Fig.~\ref{fig:sketch1p1p2}) gives the mixed-index block
\begin{equation}
\left.T^\mu{}_{\nu}\right|_{(u,\hat q)}=
\begin{pmatrix}
-\varepsilon & \ q \\[2pt]
-\,q & \ p_{\parallel}
\end{pmatrix}.
\label{eq:Tmunu_matrix}
\end{equation}
Diagonalizing this matrix requires a hyperbolic rotation that is exactly the thermodynamic boost in this plane. Under this boost, the off–diagonal component transforms as (using $T'=\Lambda T \Lambda^{-1}$ in the $(u,\hat q)$ block.)
\begin{equation}
\big(T'\big)^{0}{}_{1}=\Big(\tfrac{\varepsilon+p_{\parallel}}{2}\Big)\sinh(2\chi)+q\,\cosh(2\chi),
\end{equation}
and diagonalization sets $\big(T'\big)^{0}{}_{1}=0$. A real $\chi$ exists iff $\lvert 2q\rvert \le \varepsilon+p_{\parallel}$. The exact boost angle $\chi$ required for diagonalization is therefore
\begin{equation}
\tanh(2\chi)=\frac{2q}{\varepsilon+p_{\parallel}}.
\label{eq:boost_angle_exact}
\end{equation}

A Landau energy frame exists iff $\lvert 2q\rvert \le \varepsilon+p_{\parallel}$, equivalently the energy-flow vector in the heat plane is timelike. This is the condition that $T^{\mu}{}_{\nu}$ admit a timelike eigenvector in $\mathrm{span}\{u,\hat q\}$, ensuring a real boost angle $\chi$.

\newpage

To visualize this state space, we introduce heat-plane coordinates analogous to space-time coordinates:
\begin{align}
\mathcal{X} &\equiv \varepsilon+p_{\parallel}, \label{eq:X_coord}\\
\mathcal{T} &\equiv 2q, \label{eq:T_coord}\\
\rho &\equiv \sqrt{\mathcal{X}^{2}-\mathcal{T}^{2}}. \label{eq:rho_coord}
\end{align}
The condition $|\mathcal{T}|\le \mathcal{X}$ defines an existence cone, ensuring physicality ($2|q|\le \varepsilon+p_{\parallel}$). The boost angle in this plane defines the geometric rapidity $\eta_g=2\chi$:
\begin{equation}
\eta_g = \operatorname{artanh}\!\left(\frac{\mathcal{T}}{\mathcal{X}}\right)
= \operatorname{artanh}\!\left(\frac{2q}{\varepsilon+p_{\parallel}}\right).
\label{eq:eta_geometric}
\end{equation}
The geometric rapidity $\eta_g$ characterizes the mathematical structure of $T^{\mu\nu}$ in the state space. However, the relevant quantity for analyzing the dynamics of the fluid is the physical rapidity $\eta$, corresponding to the velocity of energy transport $v$ (Eq.~\eqref{eq:v-correct}), defined by $\eta=\operatorname{artanh}(v)$. At first order, assuming isotropic pressure ($p_{\parallel}=p$), $\eta_g \approx 2q/(\varepsilon+p)$ while $\eta\approx q/(\varepsilon+p)$, so $\eta_g \approx 2\eta$. The dynamics of the physical rapidity $\eta$ will be crucial for the Unruh analogy.

\newpage

\section{Unruh Effect in Thermodynamic Space}
\label{sec:rel-unruh}

Interpreting hydrodynamic frame transformations as hyperbolic rotations provides a geometric toolbox for dissipative relativistic fluid dynamics. The boost velocity $v$ (Eq.~\eqref{eq:v-correct}) is set by the local thermodynamic state, and as the fluid evolves along its worldline, the map to the local Landau frame changes in time. Equivalently, the fluid element experiences an effective ``acceleration'' in thermodynamic state space. Following the standard Unruh logic, where the thermal spectrum is a consequence of acceleration, we ask what spectral imprint this \emph{thermodynamic acceleration} leaves.

We define the physical thermodynamic rapidity $\eta$ corresponding to the velocity of energy transport $v$:
\begin{equation}
\eta \equiv \operatorname{artanh}(v)
= \operatorname{artanh}\!\left(\frac{q}{\varepsilon+p}\right).
\label{eq:eta-def-E}
\end{equation}
Its proper-time rate of change along the Eckart worldline defines the \emph{thermodynamic acceleration}:
\begin{equation}
a_{\rm th} \equiv u^\alpha \nabla_\alpha \eta
= u^\alpha \nabla_\alpha \operatorname{artanh}\!\left(\frac{q}{\varepsilon+p}\right).
\label{eq:ath-def-E}
\end{equation}
In an adiabatic sampling window where $a_{\rm th}$ varies slowly, the boost rapidity advances linearly with proper time:
\begin{equation}
\Delta\eta \;\simeq\; a_{\rm th}\,\Delta\tau,
\qquad
\frac{\lvert D a_{\rm th}\rvert}{a_{\rm th}^{2}} \ll 1.
\label{eq:adiabatic-map}
\end{equation}
Throughout this section, we consider a neutral, single-component fluid, work to first order in gradients with $\lvert q\rvert/(\varepsilon+p)\ll1$, and assume the adiabatic condition \eqref{eq:adiabatic-map}. For concreteness, we evaluate in the Eckart frame; to leading order, the observable $a_{\rm th}=u\!\cdot\!\nabla\eta$ is frame-independent up to $\mathcal{O}(\partial^2)$.

\begin{figure}[ht]
\centering
\includegraphics[width=0.5\textwidth]{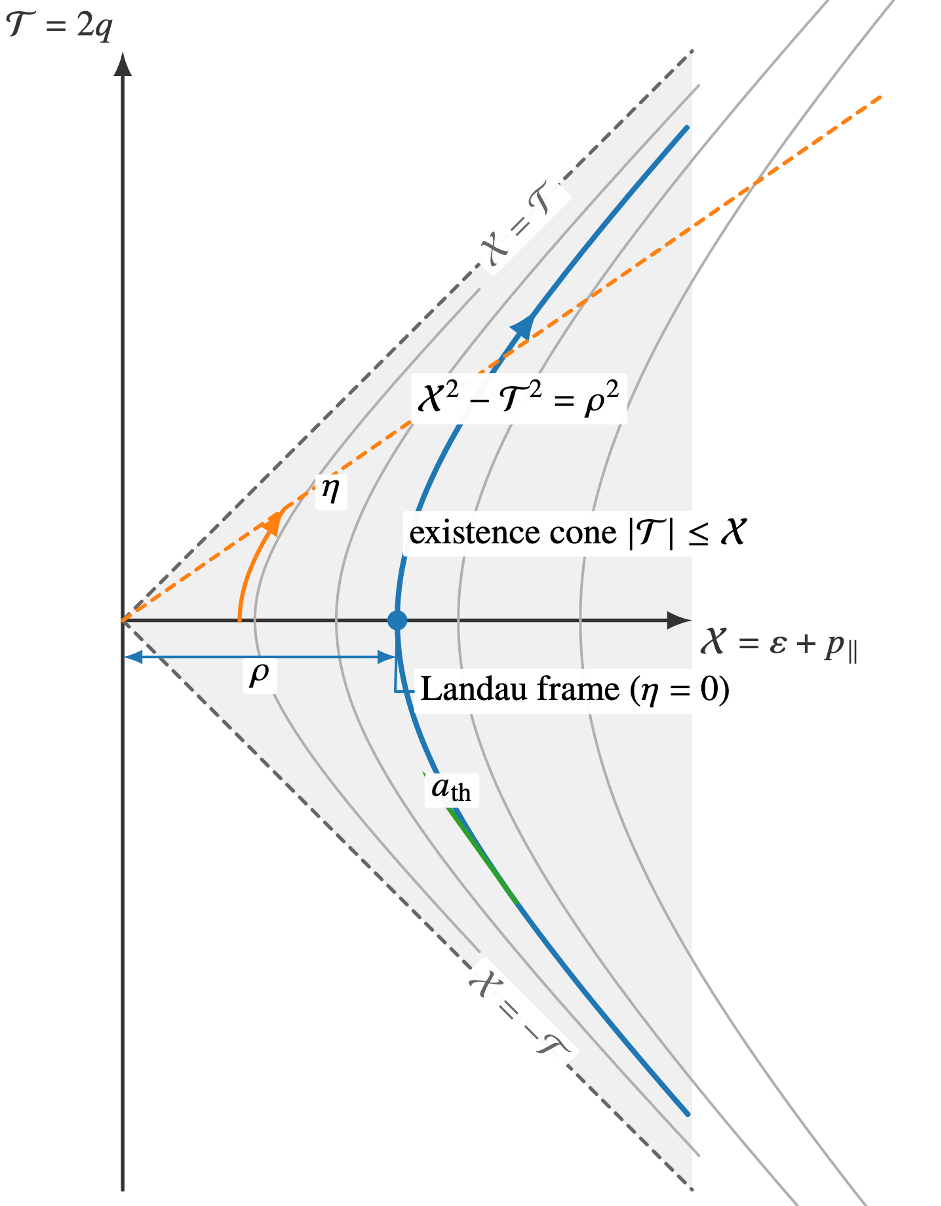}
\caption{\textbf{Thermodynamic Rindler diagram.} Visualizing the state-space geometry in the heat plane $(\mathcal X,\mathcal T)=(\varepsilon+p_{\parallel},\,2q)$. Dashed lines are the thermodynamic horizons $\mathcal X=\pm \mathcal T$. Hyperbolae $\mathcal X^{2}-\mathcal T^{2}=\rho^{2}$ are orbits under thermodynamic boosts, parameterized by the geometric rapidity $\eta_g$. Any energy frame with \(q=0\) sits at the apex (\(\eta=0\)). As the fluid evolves dynamically, its state moves in this space. The rate of change of the physical rapidity $\eta$ is the thermodynamic acceleration $a_{\rm th}=u\!\cdot\!\nabla \eta$.}
\label{fig:Rindler}
\end{figure}

In the adiabatic window, a comoving probe coupled to the heat sector sees a KMS spectrum with inverse temperature $\beta_{\rm th}=2\pi/|a_{\rm th}|$, i.e.

\begin{equation}
k_B T_{\rm th}
= \frac{\hbar\, |a_{\rm th}|}{2\pi}.
\label{eq:Tth-Unruh}
\end{equation}

\newpage

Alternatively, in QFT, the Minkowski vacuum restricted to a Rindler wedge is a KMS state with respect to the boost generator \cite{BisognanoWichmann1975,BisognanoWichmann1976, PreskillNotes}. For a stationary two-point function of a local operator $O$, the Wightman function
\begin{equation}
G^{>}(\Delta\eta)\;\equiv\;\langle O(\eta)\,O(0)\rangle
\end{equation}
is the boundary value of an analytic function in the complex $\Delta\eta$-plane satisfying the boost–KMS periodicity
\begin{equation}
G^{>}(\Delta\eta)\;=\;G^{>}(\Delta\eta+i\,2\pi).
\label{eq:KMS-boost}
\end{equation}

Using \eqref{eq:adiabatic-map} (\,$\Delta\eta=a_{\rm th}\Delta\tau$\,) this maps to a proper-time KMS relation
\begin{equation}
G^{>}(\Delta\tau)
\;=\;
G^{>}\!\left(\Delta\tau+i\,\beta_{\rm th}\right),
\qquad
\beta_{\rm th}\;\equiv\;\frac{2\pi}{a_{\rm th}}
\;=\;\frac{1}{k_B T_{\rm th}},
\label{eq:KMS-proper}
\end{equation}
which, by Fourier transform, yields the comoving KMS (detailed-balance) ratio
\begin{equation}
S_{OO}(-\omega)
\;=\;\exp\!\big[-\beta_{\rm th}\,\hbar\omega\big]\,
S_{OO}(\omega),
\label{eq:KMS-OO}
\end{equation}
with $T_{\rm th}$ given by Eq.~\eqref{eq:Tth-Unruh}. This mapping relies only on adiabaticity and standard analyticity and, to leading order, is frame-independent.

The condition in Eq.~\eqref{eq:KMS-OO} is also satisfied in modern effective field theory formulations of hydrodynamics, where the dynamical KMS symmetry is imposed as a microscopic constraint, cf.\ Refs.~\cite{CrossleyGloriosoLiu2017,HaehlLoganayagamRangamani2015,HaehlLoganayagamRangamani2018}. Hence, even without employing our geometric machinery, the existence of this symmetry within hydrodynamic EFT supports our claim and provides an alternative route to the same thermal structure. Indeed, even in the classical formulation of Landau and Lifshitz, small deviations around equilibrium are stochastic, and their variances are fixed by the fluctuation–dissipation theorem, as later emphasized in relativistic treatments.

To evaluate $a_{\rm th}$ within first-order (keeping terms through $\mathcal{O}(\partial)$ and neglecting $\mathcal{O}(\partial^{2})$), we use the approximation $\eta=v+\mathcal{O}(v^{3})$ (with $v\sim q/(\varepsilon+p)=\mathcal{O}(\partial)$). Hence,
\begin{align}
a_{\rm th} &\simeq
u_{(E)}^\alpha \nabla_\alpha\!\left(\frac{q_E}{\varepsilon_E+p_E}\right)
\label{eq:ath-linear-expansion-E}\\
&= \frac{1}{\varepsilon_E+p_E}\,u_{(E)}^\alpha \nabla_\alpha q_E
- \frac{q_E}{(\varepsilon_E+p_E)^2}\,
u_{(E)}^\alpha \nabla_\alpha(\varepsilon_E+p_E).\nonumber
\end{align}
With the hydrodynamic counting $q_E\sim\mathcal{O}(\partial)$, the second term involves products of gradients and is thus $\mathcal{O}(\partial^{2})$. Here $q_E=\mathcal{O}(\partial)$ and $u\!\cdot\!\nabla(\varepsilon_E+p_E)=\mathcal{O}(\partial)$, so their product is $\mathcal{O}(\partial^{2})$. Neglecting this higher-order term gives
\begin{equation}
a_{\rm th} \simeq \frac{1}{\varepsilon_E+p_E}\, u_{(E)}^\alpha \nabla_\alpha q_E.
\label{eq:ath-linear-E}
\end{equation}
The thermodynamic acceleration is primarily driven by the proper-time derivative of the heat flux.

To express $a_{\rm th}$ in terms of standard hydrodynamic variables and transport coefficients, we employ the causal Israel--Stewart (Maxwell--Cattaneo) relaxation equation:
\begin{equation}
\tau_q\, u_{(E)}^\alpha \nabla_\alpha q^{\langle \mu\rangle} + q^\mu
\simeq -\kappa \big(\nabla^\mu T + T\, a_{(E)}^\mu\big),
\label{eq:IS-equation-E}
\end{equation}
where $\tau_q$ is the relaxation time, $\kappa$ is the thermal \emph{conductivity}, $T$ is the temperature, and $a_{(E)}^\mu\equiv u_{(E)}^\alpha \nabla_\alpha u_{(E)}^\mu$ is the fluid acceleration in space-time. Projecting this equation along the heat direction ($\nabla_{\parallel}\equiv \hat q^{\mu}\nabla_{\mu}$, $a_{\parallel}\equiv \hat q^{\mu} a_{(E)\mu}$) gives
\begin{equation}
u_{(E)}^\alpha \nabla_\alpha q_E
\simeq -\frac{1}{\tau_q}\Big[q_E + \kappa\big(\nabla_\parallel T + T\, a_\parallel\big)\Big].
\label{eq:Dq-proj-E}
\end{equation}
Substituting this into Eq.~\eqref{eq:ath-linear-E} yields the explicit form of the thermodynamic acceleration:
\begin{equation}
a_{\rm th} \simeq -\frac{1}{\tau_q\,(\varepsilon_E+p_E)}
\Big[q_E + \kappa\big(\nabla_\parallel T + T\, a_\parallel\big)\Big].
\label{eq:ath-hydro-E}
\end{equation}
Therefore, the predicted Thermodynamic Unruh temperature is
\begin{equation}
k_B T_{\rm th} \simeq \frac{\hbar}{2\pi\,\tau_q\,(\varepsilon_E+p_E)}
\left| q_E + \kappa\big(\nabla_\parallel T + T\, a_\parallel\big)\right|.
\label{eq:Tth-hydro-E}
\end{equation}

It is important to note that $T_{\rm th}$ is distinct from the material temperature $T$. The temperature $T$ determines the overall amplitude of fluctuations through fluctuation–dissipation relations. By contrast, $T_{\rm th}$ is the Kubo--Martin--Schwinger (KMS) temperature. This temperature represents the balance ratio of comoving fluctuations, as the probe samples fluctuations using non-inertial time associated with the rapidity change $\eta$. Hence, the Thermodynamic Unruh effect is a universal indicator of transient, out-of-equilibrium dynamics in first-order causal hydrodynamics. The issues of stability in Israel--Stewart framework \cite{HiscockLindblom1983} and other modern theories discussed in subsequent sections, are inherent to the theory at hand and do not affect the result.
\newpage

\section{Universality of the Effect}
\label{sec:universality}

The Thermodynamic Unruh signal is based on a single symmetry-fixed driver of the system. Thus, it does not depend on a closure scheme or a specific hydrodynamic frame if we stay within a neutral, single-component fluid (no conserved charge, no external electromagnetic fields) and work to first order in gradients. In this setting, the leading covariant thermal driving force is
\begin{equation}
E_T^\mu \;\equiv\; \Delta^{\mu\nu}\!\left(\nabla_\nu T + T\,a_\nu\right),
\label{eq:ET-def-main}
\end{equation}
which is orthogonal to $u^\mu$ by construction. Tolman--Ehrenfest equilibrium is the \emph{four-vector} condition\cite{Tolman1930,TolmanEhrenfest1930,LandauLifshitz1987}
\begin{equation}
\nabla_\mu T + T a_\mu = 0,
\label{eq:tolman}
\end{equation}
implying $E_T^\mu=0$ and $u\!\cdot\!\nabla T=0$. Geometrically, one may regard $E_T^\mu$ as the direction in space-time toward which the system is thermally “pulled”; in the heat plane, this vector governs the motion of the \emph{thermodynamic boost} rapidity $\eta$. The observable we propose is the comoving KMS ratio with effective temperature $T_{\rm th}$ defined in Eq.~\eqref{eq:Tth-Unruh}, which is therefore tied to the thermodynamic acceleration of Eq.~\eqref{eq:ath-hydro-E} through the evolution of $E_T^\mu$ and $q^\mu$ along the worldline.

In modern general-frame first-order theories (BDNK)\cite{BemficaPRX2022,GavassinoPRX2022}, the hydrodynamic variables themselves include $\mathcal{O}(\partial)$ redefinitions, so the frame is effectively dynamical; nevertheless, the leading-order observable $a_{\rm th}=u\!\cdot\!\nabla\eta$ is unchanged. (see Appendix~\ref{sec:relfluid} for a review of these frameworks),

A causal hyperbolic realization in the Eckart frame follows from the Israel--Stewart relaxation equation \eqref{eq:IS-equation-E} and, with manipulations, yields Eq.~\eqref{eq:Tth-hydro-E}. Two corollaries at this order are:
\begin{enumerate}\item[(i)] In Tolman--Ehrenfest equilibrium ($E_T^\mu=0$), one has $q_E\to0$, so $a_{\rm th}\to0$ and the signal vanishes even at finite material temperature $T$.
\item[(ii)] In non-equilibrium steady states with $u\!\cdot\!\nabla q_E\simeq0$, Eq.~\eqref{eq:ath-linear-E} gives $a_{\rm th}\simeq0$.
\end{enumerate}
Thus, the signal diagnoses transient out-of-equilibrium dynamics.

Although the above expressions were written in the Eckart frame, the out-of-equilibrium frame is not unique, and there are infinitely many equivalent options. Under admissible first-order frame redefinitions, $u^\mu$ and $q^\mu$ shift by $\mathcal O(\partial)$, so $\eta=\operatorname{artanh}\!\big(q/(\varepsilon+p)\big)=\mathcal O(\partial)$ shifts by $\mathcal O(\partial)$, and $a_{\rm th}=u\!\cdot\!\nabla\eta$ changes only at $\mathcal O(\partial^2)$. Consequently, \emph{to the order we work}, the observable (the comoving KMS ratio and the inferred $T_{\rm th}$) is representation-independent, provided the detector coupling is transformed consistently with the fields (see Refs.~\cite{Kovtun2019}
 for general-frame discussions). Thus, changing frames is akin to changing coordinates: components shift, but the invariant content of the signal remains unchanged.

If we dispense with an explicit relaxation variable as in pure first-order, general-frame hydrodynamics, the heat flux is algebraic at leading order, $q^\mu=-\kappa\,E_T^\mu+\cdots$. Then
\begingroup
\setlength{\abovedisplayskip}{6pt}
\setlength{\belowdisplayskip}{6pt}
\begin{equation}
u\!\cdot\!\nabla q_E=-\kappa\,u\!\cdot\!\nabla E_{T,\parallel}+\cdots
\end{equation}
and Eq.~\eqref{eq:ath-linear-E} gives
\begin{equation}
a_{\rm th}\;\simeq\; -\,\frac{\kappa}{\varepsilon_E+p_E}\,
u_{(E)}^\alpha \nabla_\alpha\!\big(\nabla_\parallel T + T a_\parallel\big)
\;+\; \mathcal O(\partial^3),
\label{eq:ath-pure-first-order}
\end{equation}
\endgroup
So the effect appears at $\mathcal O(\partial^2)$ and is nonzero whenever the generalized thermal force varies along the worldline during a transient.

In both the relaxation-dominated and pure first-order regimes, the mapping in Eq.~\eqref{eq:Tth-Unruh} establishes a unified definition of $T_{\rm th}$ based on the comoving Kubo--Martin--Schwinger (KMS) ratio within an adiabatic sampling window, where $a_{\rm th}$ is approximately constant. 

Consequently, the Thermodynamic Unruh effect is a universal consequence of the discussed heat-plane kinematics determined by the symmetry-fixed $E_T^\mu$ in Eq.~\eqref{eq:ET-def-main}.

In this section, we assumed a neutral, single-component fluid without external electromagnetic fields and neglected parity-odd and vortical couplings. It is important to note that additional first-order contributions may arise from sources such as conserved charges, electromagnetic fields, strong rotational effects, or higher-order transport phenomena. Hence, these factors would alter the expression for $a_{\rm th}$ nevertheless, the fundamental scaling with $\lvert u\!\cdot\!\nabla\eta\rvert$ remains unchanged

\section{Conclusion}
\label{sec:conclusion}

In this work, we have established a geometric analogy between the structure of relativistic hydrodynamics and the kinematics underlying the Unruh effect. We demonstrated that the fundamental ambiguity in defining the local rest first-order frame, as in the distinction between the Eckart and Landau--Lifshitz frames, can be interpreted as a local hyperbolic rotation whose instantaneous map is a Lorentz-like boost (Eqs.~\eqref{eq:landau-boost-eps}--\eqref{eq:landau-boost-q}) operating within an abstract thermodynamic state space. Then, when we consider the time derivative of this "Thermodynamic Boost", or equivalently, the thermodynamic acceleration, $a_{\rm th}$,(Eq.~\eqref{eq:ath-def-E}), its source being the hyperbolic rotation between the frames, results in the relativistic fluid counterpart to the quantum Bogoliubov transformation.

Our central prediction in this paper is that this thermodynamic acceleration  inherits from its quantum counterpart a distinct spectral signature, the Thermodynamic Unruh temperature, $T_{\rm th}$ (Eq.~\eqref{eq:Tth-Unruh}). Then, using the causal framework of Israel--Stewart theory, we derived an expression for this temperature (Eq.~\eqref{eq:Tth-hydro-E}), which is a universal feature of transient, out-of-equilibrium systems in causal first-order relativistic hydrodynamics.

Hence,  in the presence of dissipative processes, relativistic fluids do not acquire an extra thermal bath in the usual sense; rather, during transients, a comoving heat–sector probe exhibits a KMS spectral asymmetry that can be parameterized by a Thermodynamic Unruh temperature.

\begin{equation}
T_{\rm th}
= \frac{\hbar}{2\pi k_B}\,\big|u\!\cdot\!\nabla\eta\big|
\;\simeq\;
\frac{\hbar}{2\pi k_B\,\tau_q(\varepsilon+p)}
\left|\, q + \kappa\big(\nabla_{\!\parallel}T + T\,a_{\parallel}\big)\right|.
\label{eq:Tth-summary}
\end{equation}

Hence, this additional thermal signature $T_{\rm th}$ is \emph{distinct from} the material temperature $T$ and vanishes in Tolman--Ehrenfest balance ($\nabla T+Ta=0$) or steady conduction ($u\!\cdot\!\nabla q\simeq 0$), but it is \emph{generically nonzero} whenever the thermodynamic boost changes in time (equivalently, $u\!\cdot\!\nabla\eta\neq 0$).

Since it is unreasonable to expect direct experimental confirmation of this result, we can investigate it through simulations. Although the BDNK theory is the more comprehensive theory available, the Israel--Stewart theory is the one most commonly used in simulations.\cite{MostNoronha2021} For space–plasma/GRMHD–like conditions (a weakly dissipative fluid), the heat–to–enthalpy ratio is conservatively in the $10^{-3}$–$10^{-2}$ range and the relaxation time in the $5$–$50~{\rm ms}$ band (we use Eqs.~\eqref{eq:IS-equation-E}, \eqref{eq:ath-hydro-E}–\eqref{eq:Tth-hydro-E}). This gives
\begin{equation}
T_{\rm th}\;\sim\;10^{-13}\text{--}10^{-11}\ {\rm K},\qquad
f_\ast \;\equiv\; \frac{k_B T_{\rm th}}{h}\;\sim\;{\rm mHz}\text{--}0.1~{\rm Hz},
\label{eq:Tth-fstar-weak-range}
\end{equation}
which is small, well below instrumental bands, but present. By contrast, in \emph{strong, ultrafast transients} (a far–from–equilibrium state) with a rapid, microscopic relaxation time and an order–one drive, one expects
\begin{equation}
T_{\rm th}\;\sim\;{\rm mK}\text{--}{\rm K},\qquad
f_\ast\;\sim\;{\rm MHz}\text{--}{\rm GHz},
\label{eq:Tth-fstar-strong-range}
\end{equation}
making the KMS slope observable.

Lastly, we extended our central argument to modern relativistic fluid frameworks to demonstrate, irrespective of closure or frame choice, an Unruh–type spectral signal persists across first-order hydrodynamic frames, cf.\ Eq.~\eqref{eq:ath-pure-first-order} characterized by the Thermodynamic Unruh temperature $T_{\rm th}$.

With the universality of the effect on one hand and an observable prediction on the other, we show that the rich structure of relativistic dissipative fluids can mimic and reproduce quantum phenomena, casting a classical shadow of observer-dependent quantum fields. Hinting at a broader pattern, this correspondence might be the dawn of a deeper framework.

\section*{Acknowledgments}

I thank Prof.~Elias Most for his inspiring talk and his feedback, which encouraged me to explore these ideas. I also thank Prof.~Ulf Leonhardt for his mentorship, companionship, and for teaching me the intricacies of the Unruh effect. Additionally, I would like to thank Adem Deniz Pişkin for our fruitful discussions. Finally, I thank Prof.~Saul Teukolsky for accepting me as a visiting researcher at TAPIR, Caltech, where the first draft of this work was written. I gratefully acknowledge the support of the Weizmann Institute of Science.

\begin{center}
\textbf{\large Appendix}
\end{center}

\setcounter{section}{0}
\renewcommand{\thesection}{\Alph{section}}

\setcounter{equation}{0}
\makeatletter
\@addtoreset{equation}{section}
\makeatother
\renewcommand{\theequation}{\thesection\arabic{equation}}

\setcounter{figure}{0}
\makeatletter
\@addtoreset{figure}{section}
\makeatother
\renewcommand{\thefigure}{\thesection\arabic{figure}}

\setcounter{table}{0}
\makeatletter
\@addtoreset{table}{section}
\makeatother
\renewcommand{\thetable}{\thesection\arabic{table}}

\section{Relativistic Fluid Mechanics}
\label{sec:relfluid}

The principle of relativity has implications for the bulk masses studied in classical mechanics, as well as for the propagation of any form of signal, including heat flux and viscous stresses. Hence, all equations in physics, irrespective of the domain, need to harbor non-trivial relativistic corrections.
Even at nonrelativistic velocities, the classical transport equations, such as Fourier’s heat equation, fail to respect relativistic causality. Hence, classical heat conduction, described by the diffusion equation
\begin{equation}
\partial_t T - D_T\nabla^{2}T = 0,
\label{eq:diffusion}
\end{equation}
where the thermal diffusivity is
\begin{equation}
D_T \;=\; \frac{\kappa}{\varrho\,c_p},
\label{eq:kappa-def}
\end{equation}
with $\kappa$ the thermal conductivity, $\varrho$ the mass density, and $c_p$ the specific heat capacity.
\begin{figure}[ht]
\centering
\includegraphics[width=0.8\textwidth]{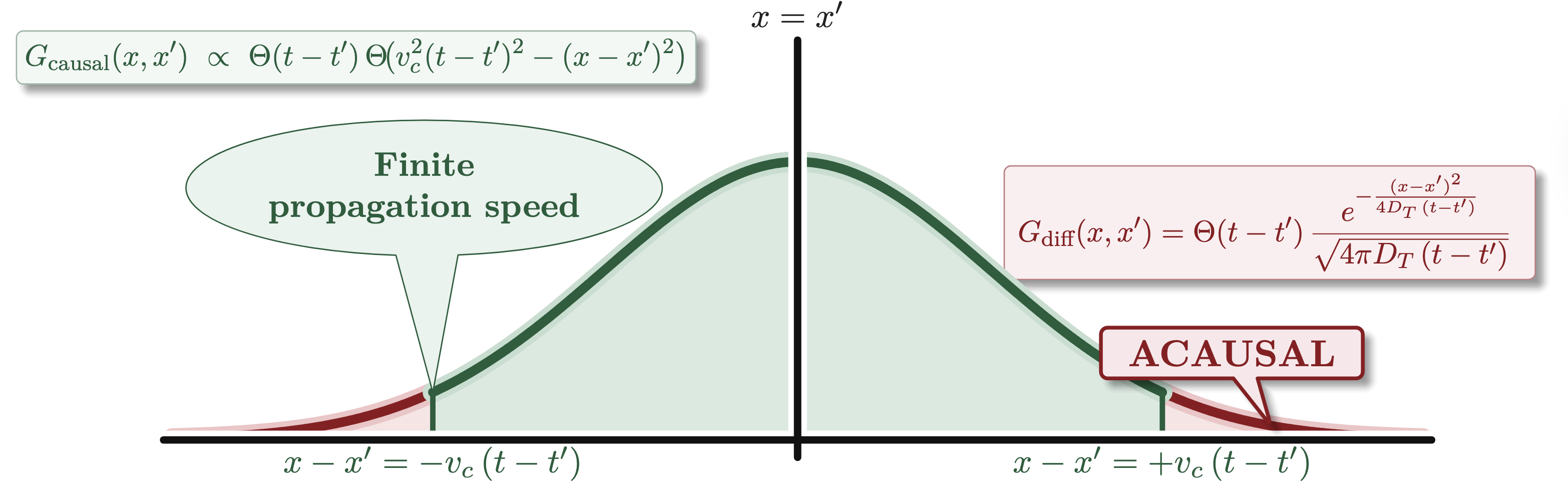}
\caption{\textbf{Causal vs.\ acausal heat propagation.}
The red curve illustrates the diffusion response, characterized by instantaneous spatial tails that extend beyond the light cone. The Green function extending to infinity corresponds to the acausal character. The relativistic correction is illustrated by the Green curve, which results in a more complex Green function.}
\label{fig:GreenFunc}
\end{figure}

The diffusion equation is a parabolic PDE and predicts instantaneous propagation of thermal disturbances, thereby violating causality. Making relativistic correction corresponds to the addition of a relaxation term, resulting in the hyperbolic telegrapher's equation: \footnote{The equation was first discovered by transmission line engineers, where in this domain the relaxation time corresponds to the delay due to inductive elements}
\begin{equation}
\tau\,\partial_t^{2}T + \partial_t T - D_T\nabla^{2}T = 0,
\label{eq:telegraph}
\end{equation}
whose characteristic propagation speed
\begin{equation}
v_c=\sqrt{\frac{D_T}{\tau}},
\label{eq:vc}
\end{equation}
is constrained to satisfy $v_c\leq c$. The relaxation time $\tau$ thus ensures causal consistency within the theory \cite{Cattaneo1958,Vernotte1958}. Hence, just as the heat equation requires modification, the theory of fluids encompassing the Navier–Stokes equations and the associated conserved quantities must also admit relativistic corrections.
A fully covariant treatment of relativistic fluids begins with the stress--energy tensor decomposed with respect to a chosen fluid four-velocity $u^{\mu}$:
\begin{equation}
T^{\mu\nu}= \varepsilon u^{\mu}u^{\nu}
+ p\,\Delta^{\mu\nu}
+ q^{\mu}u^{\nu}+q^{\nu}u^{\mu}
+ \pi^{\mu\nu}, \quad
\Delta^{\mu\nu}=g^{\mu\nu}+u^{\mu}u^{\nu},
\label{eq:tmunu-decomp}
\end{equation}
where $\varepsilon$ is the energy density, $p$ is the (isotropic) pressure, $q^{\mu}$ the heat flux, and $\pi^{\mu\nu}$ the shear stress tensor, both orthogonal to $u^{\mu}$.
The shear stress tensor $\pi^{\mu\nu}$ describes anisotropic stresses, unlike pressure, which captures isotropic stresses. It is symmetric $\pi^{\mu\nu} = \pi^{\nu\mu}$, traceless $\pi^{\mu}{}_{\mu} = 0$, and by definition orthogonal to the fluid velocity $\pi^{\mu\nu} u_\nu = 0$.
In relativistic fluid theory, the decomposition into dissipative and non-dissipative parts is not unique, but depends on the choice of fluid frame. At the epoch of relativistic fluid dynamics, two choices were present:
\emph{Eckart (particle) frame}: The four-velocity $u^{\mu}$ is parallel to the conserved particle current $n^{\mu}$. In this frame, particle diffusion vanishes identically, while a heat flux $q^{\mu}$ generally appears.
\emph{Landau--Lifshitz (energy) frame}: The four-velocity $u^{\mu}$ is defined as the timelike eigenvector of the stress--energy tensor \(T^{\mu\nu}\)
. By construction, this frame eliminates energy flux orthogonal to $u^{\mu}$, setting $q^{\mu}=0$, while a corresponding particle diffusion current emerges.
Switching between these two frames is accomplished to the first order by
\begin{equation}
u^{\mu}_{(L)} =
u^{\mu}_{(E)} +
\frac{q^{\mu}_{(E)}}{\varepsilon+p}
+ \mathcal O(q^{2}).
\label{eq:uL-linear}
\end{equation}

This suggests that the observed heat and particle currents in relativistic fluid dynamics are observer-dependent quantities, depending upon the choice of hydrodynamic frame. 
The transformation between these two frames and its dependence on the choice of frame is the crucial observation for the purposes of this paper. However, at first order, the Eckart and Landau frames are ill-posed and acausal frameworks, motivating the next generation of physicists to investigate further.
For satisfying covariance and causality simultaneously, Israel and Stewart developed a framework where dissipative fluxes $q^{\mu}$ and $\pi^{\mu\nu}$ are treated as independent dynamical variables obeying relaxation-type equations similar to the correction to the Fourier law we discussed, see also \cite{Muller1967}:
\begin{align}
\tau_q\,u^{\alpha}\nabla_{\alpha} q^{\langle\mu\rangle}+q^{\mu}
&= -\kappa\left(\nabla^{\mu}T + T a^{\mu}\right),\nonumber\\
\tau_\pi\,u^{\alpha}\nabla_{\alpha}\pi^{\langle\mu\nu\rangle}+\pi^{\mu\nu}
&= 2\eta_{\mathrm{s}}\,\sigma^{\mu\nu},
\label{eq:IS-eqs}
\end{align}
where $a^{\mu}=u^{\alpha}\nabla_{\alpha}u^{\mu}$ denotes the fluid acceleration and $\sigma^{\mu\nu}$ is the shear tensor. These equations, hyperbolic by construction, enforce finite signal propagation speeds and stability.
However, this causal framework comes with a subtlety: since it constitutes an effective theory truncated at first order in gradients, the entropy production rate,
\begin{equation}
\nabla_{\mu}S^{\mu}
= \frac{q^{\mu}q_{\mu}}{\kappa T^{2}} + \frac{\pi^{\mu\nu}\pi_{\mu\nu}}{2\eta_{\mathrm{s}} T} + \dots,
\label{eq:entropy}
\end{equation}
remains strictly positive only in regimes where dissipative fluxes are small compared to background thermodynamic quantities ($q^{\mu},\pi^{\mu\nu}\ll \varepsilon$). Thus, large gradients beyond the first term naturally indicate the breakdown of the first-order effective description and naturally necessitate higher-order theories or microscopic kinetic treatments.

Ultimately, frame choice and hyperbolic relaxation are not optional adjustments, but are fundamental features enforced by relativistic causality.

While the Müller--Israel--Stewart (MIS) program provided the first systematic approach to addressing the acausality and instability of first-order relativistic Navier–Stokes equations, it is not the final verdict on resolving these issues.
Subsequent advances have established a robust conceptual foundation from four complementary perspectives. 

First is symmetry considerations, as formalized by BRSSS, which enumerate the admissible second-order tensors and their interrelations that clarify which of the permitted structures are permitted or excluded by symmetry~\cite{BRSSS2008}. 
Second is the kinetic theoretical perspective, as developed in DNMR, that derives transient hydrodynamics from the Boltzmann equation using systematic moments, demonstrating the generic nature of relaxation terms and relating relaxation times to microscopic scales~\cite{DNMR2012}. 

The third approach is to recast relativistic fluid dynamics as an effective field theory (EFT), which imposes constraints. Enforcing dynamical KMS symmetry incorporates the second law of thermodynamics and fluctuation–dissipation relations, and identifies the generalized thermal driving four–vector.
The square of this vector determines the thermal contribution to entropy production. 
The fourth and the most recent contribution of the BDNK is the use of general frames to resolve causality at first order: by permitting controlled $\mathcal O(\partial)$ redefinitions of $(T,u^\mu)$, it is possible to formulate symmetric–hyperbolic, subluminal conservation laws. This demonstrates that the acausality associated with Landau/Eckart matching arises from a choice of frame rather than representing a fundamental limitation.

The fundamental basis of our discussion has two elements from the elements of modern relativistic fluid dynamics:
\begin{enumerate}[label=(\roman*), topsep=0pt, itemsep=0pt, parsep=0pt, partopsep=0pt]
\item \emph{The symmetry-fixed driver \(E_T^\mu\).} As we also discussed in the text
\begin{equation}
E_T^{\mu}\;\equiv\;\Delta^{\mu\nu}\!\left(\nabla_{\nu}T+T\,a_{\nu}\right),
\label{eq:ET-app}
\end{equation}
The four-vector \(E_T^\mu\) in Eq.~\eqref{eq:ET-app} determines the unique first-order thermal driving direction. When \(E_T^\mu=0\) (Tolman–Ehrenfest balance), the relevant mechanism is absent. When \(E_T^\mu\neq0\), no alternative first-order driver is present in the analysis. Higher-order frameworks such as BRSSS and EFT are relevant only to modify \(E_T^\mu\) with subleading gradients, hence they do not introduce a new \(\mathcal O(\partial)\) direction, and therefore the conclusion stands the same.
\item \emph{A causal evolution law for the heat sector.} One may close algebraically at leading order,
\begin{equation}
q^\mu \;=\; -\,\kappa\,E_T^\mu \;+\;\mathcal O(\partial^{2}),
\label{eq:q-alg-app}
\end{equation}
or include relaxation in the Israel--Stewart/Maxwell--Cattaneo form (Eq.~\eqref{eq:IS-eqs}). This choice serves as a gauge for how quickly the system responds and which coefficients multiply the same kinematic content; hence, it similarly does not alter that content. To see mathematically in both cases, the relevant small parameter
\begin{equation}
v \;\sim\; \frac{q}{\varepsilon+p}\;=\;\mathcal O(\partial),
\end{equation}
that defines the physical rapidity \(\eta=\operatorname{artanh}(v)\). During transient processes, the worldline derivative \(u\!\cdot\!\nabla\eta\) may be nonzero. It is the effective acceleration at this order that is the only scalar obtained from the heat sector that is invariant under frame redefinitions. Hence, general-frame shifts only relabel \((T,u^\mu)\) at \(\mathcal O(\partial)\) that alter the coefficients, but the existence of this driver remains intact.
\end{enumerate}

\newpage
\section{Unruh Effect}
\label{sec:unruh}

The essence of the Rindler diagram is key to many aspects of relativity~\cite{Leonhardt2010}. At its core, it represents the consequences of acceleration in space-time diagrams. Combining with the equivalence principle, the Rindler diagram offers the most intuitive lift from SR to GR. Its boundaries are equivalent to the event horizon of a black hole, where the central discontinuity between wedges is reminiscent of that of a wormhole. So, it is no surprise that, like other problems at the frontiers of physics, the Unruh effect also boils down to a Rindler diagram~\cite{Rindler1966}.

\begin{figure}[h]
\centering
\includegraphics[width=0.4\textwidth]{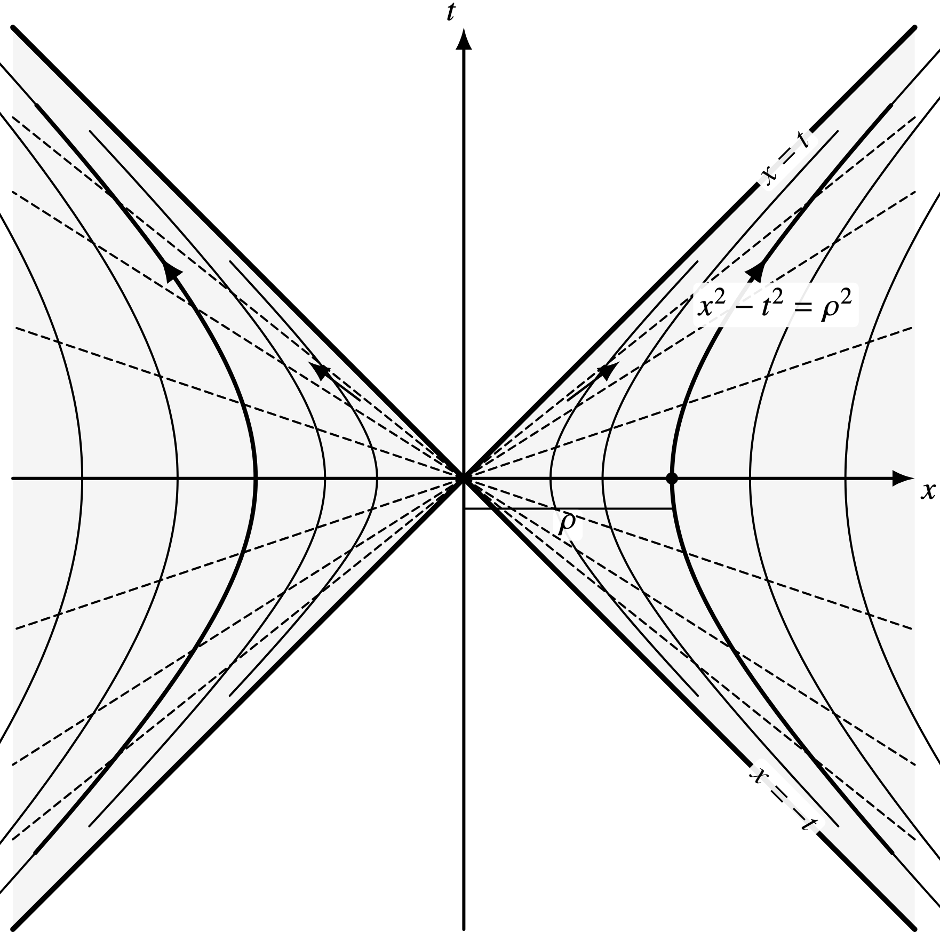}
\caption{\textbf{Rindler diagram.}
Rindler coordinates as in \eqref{eq:rindler-transform}:
\(x=\xi\cosh(a\eta_{\mathrm R}/c),\; ct=\xi\sinh(a\eta_{\mathrm R}/c)\).
Solid curves are \(\xi=\text{const}\) (uniform acceleration, \(a=c^{2}/|\xi|\));
dashed lines are \(\eta_{\mathrm R}=\text{const}\).
The null lines \(x=\pm ct\) bound the right/left wedges (Rindler horizons).
Along \(\xi=\text{const}\), \(d\tau=(a\xi/c^{2})\,d\eta_{\mathrm R}\).}
\label{fig:RindlerClassical}
\end{figure}

The worldline of a uniformly accelerated observer (proper acceleration \(a\)) in an inertial (Minkowski) frame is the hyperbola
\begin{equation}
x^{2}-c^{2}t^{2}=\left(\frac{c^{2}}{a}\right)^{2},
\label{eq:hyperbola}
\end{equation}
which lies in the right or left Rindler wedges. Adopting hyperbolic (Rindler) coordinates \((\eta_{\mathrm R},\xi)\),
\begin{equation}
x=\xi\,\cosh\!\left(\frac{a\eta_{\mathrm R}}{c}\right),\qquad
ct=\xi\,\sinh\!\left(\frac{a\eta_{\mathrm R}}{c}\right),
\label{eq:rindler-transform}
\end{equation}
The Minkowski metric becomes
\begin{equation}
ds^{2}
= -\left(\frac{a\xi}{c}\right)^{2} d\eta_{\mathrm R}^{2}
  + d\xi^{2}+dy^{2}+dz^{2}.
\label{eq:rindler-metric-c}
\end{equation}
Curves of constant \(\xi\) are precisely the hyperbolae \eqref{eq:hyperbola}; each has proper acceleration
\begin{equation}
a(\xi)=\frac{c^{2}}{\xi},\qquad (\xi>0),
\label{eq:proper-accel}
\end{equation}
so the trajectory with \(\xi=c^{2}/a\) has proper acceleration \(a\).
Lines of constant \(\eta_{\mathrm R}\) are straight rays through the origin,
\begin{equation}
\frac{ct}{x}=\tanh\!\left(\frac{a\eta_{\mathrm R}}{c}\right),
\label{eq:constant-eta}
\end{equation}
and the surface \(\xi=0\) (equivalently \(|x|=ct\)) is the Rindler horizon.
From \eqref{eq:rindler-metric-c}, along \(\xi=\mathrm{const}\) one has
\begin{equation}
d\tau=\frac{a\xi}{c^{2}}\,d\eta_{\mathrm R} .
\label{eq:proper-time-reln}
\end{equation}

Quantizing a free field with respect to the Minkowski time flow \(\partial_t\) or the Rindler time flow \(\partial_{\eta_{\mathrm R}}\)
produces two complete positive–frequency bases
\(\{u_k,u_k^*\}\) and \(\{v_\omega,v_\omega^*\}\) with corresponding ladder operators
\(\{\hat a_k,\hat a_k^\dagger\}\) and \(\{\hat b_\omega,\hat b_\omega^\dagger\}\).
The field admits both decompositions,
\begin{equation}
\hat\phi(x)=\int_{-\infty}^{\infty}\!{\rm d}k\;\big(u_k \hat a_k + u_k^* \hat a_k^\dagger\big)
           =\int_{0}^{\infty}\!{\rm d}\omega\;\big(v_\omega \hat b_\omega + v_\omega^* \hat b_\omega^\dagger\big).
\label{eq:two-decomps}
\end{equation}
(see ~\cite{BirrellDavies1982,Wald1994,ParkerToms2009}).
Completeness yields the Bogoliubov relations for modes and operators,
\begin{align}
u_k &= \int_{0}^{\infty}\!{\rm d}\omega\;\Big(\alpha_{\omega k}\,v_\omega + \beta_{\omega k}\,v_\omega^{*}\Big), 
\label{eq:mode-bogo}
\\[2pt]
\hat b_\omega &= \int_{-\infty}^{\infty}\!{\rm d}k\;\Big(\alpha_{\omega k}\,\hat a_k + \beta_{\omega k}\,\hat a_k^\dagger\Big),
\label{eq:op-bogo}
\end{align}
with the canonical algebra enforcing
\begin{align}
\int_{-\infty}^{\infty}\!{\rm d}k\;
\big(\alpha_{\omega k}\alpha_{\omega' k}^{*}-\beta_{\omega k}\beta_{\omega' k}^{*}\big)
&= \delta(\omega-\omega'),
\label{eq:bogo-constraint-1}
\\
\int_{0}^{\infty}\!{\rm d}\omega\;
\big(\alpha_{\omega k}\alpha_{\omega k'}^{*}-\beta_{\omega k}\beta_{\omega k'}^{*}\big)
&= \delta(k-k').
\label{eq:bogo-constraint-2}
\end{align}
Proper, orthochronous Lorentz transformations preserve the sign of frequency for plane waves
\(u_k(x)=e^{-ik\!\cdot\!x}\) and therefore do not create particles:
\begin{equation}
\beta=0\quad\text{under proper Lorentz transformations}.
\label{eq:beta-zero-Lorentz}
\end{equation}
\cite{PreskillNotes}
Nonzero \(\beta\) arises only when comparing distinct timelike Killing flows.
Passing from \(\partial_t\) (Minkowski time) to \(\partial_{\eta_{\mathrm R}}\) (Rindler time) mixes positive and
negative Minkowski frequencies across the horizon, yielding a thermal spectrum.

Let \(\lvert 0_M\rangle\) be the Minkowski vacuum annihilated by every \(\hat a_k\).
For the Rindler number operator \(\hat N_\omega=\hat b_\omega^\dagger \hat b_\omega\),
\begin{equation}
\langle 0_M | \hat N_\omega | 0_M \rangle
= \int_{-\infty}^{\infty}\!{\rm d}k\; \big|\beta_{\omega k}\big|^{2}
= \frac{1}{\exp\!\big(2\pi\omega/a\big)-1},
\label{eq:planck-N}
\end{equation}
which is thermal at the Unruh temperature
\begin{equation}
T_{\mathrm U}
= \frac{\hbar\,a}{2\pi c\,k_B}
\qquad \big(c=\hbar=k_B=1 \Rightarrow T_{\mathrm U}=a/2\pi\big).
\label{eq:unruh-T}
\end{equation}

Equivalently, as discussed in the main text, this thermal spectrum can be encoded in the statement that the two-point function along the accelerated worldline satisfies the Kubo–Martin–Schwinger (KMS) periodicity.

The Minkowski vacuum can be thought of as a two–mode squeezed state entangling right (\(\mathcal{R}\)) and left (\(\mathcal{L}\)) wedges:
\begin{equation}
\lvert 0_M\rangle
=\prod_{\omega}\!\left[ \sqrt{1-e^{-2\pi\omega/a}}
\sum_{n=0}^{\infty} e^{-n\pi\omega/a}\,
\lvert n_\omega\rangle_{\mathcal{R}}\!\otimes\!\lvert n_\omega\rangle_{\mathcal{L}} \right].
\label{eq:squeezed}
\end{equation}
Tracing over \(\mathcal{L}\) yields, for each mode, the thermal state
\begin{equation}
\rho_{\mathcal{R},\omega}=(1-e^{-2\pi\omega/a})
\sum_{n=0}^{\infty} e^{-2\pi\omega n/a}\,
\lvert n_\omega\rangle_{\mathcal{R}}\!\langle n_\omega\rvert ,
\end{equation}
whose occupation numbers match \eqref{eq:planck-N} at temperature \(T_{\mathrm U}\).
Restricting the global vacuum state to a single Rindler wedge, by tracing out the degrees of freedom in the opposite wedge, results in a mixed thermal state. This leads to an apparent loss of information for observers experiencing the Unruh temperature $T_{\mathrm {U}} $, even though the global quantum state remains pure. Hence, the Unruh effect can be interpreted as a measure of this mixing relative to the Minkowski vacuum.

\end{document}